\documentstyle[11pt,aaspp4]{article}
\input{psfig}
\eqsecnum
\begin{document}
\title{Lithium Production in Hot Advection-Dominated Accretion Flows in
Soft X-Ray Transients}

\author{Insu Yi and Ramesh Narayan\footnote{Permanent address: 
Harvard-Smithsonian Center for Astrophysics, 60 Garden St., Cambridge,
MA 02138}}
\affil{Institute for Advanced Study, Olden Lane, Princeton, NJ 08540}


\begin{abstract}
High Li abundances have been reported in the late type secondaries of
five soft X-ray transients (SXTs), V404 Cyg, A0620-00, GS2000+25, Nova
Mus 1991, and Cen X-4.
Since Li is likely to be depleted in stars of this type, the
origin of the Li is puzzling.  Li has not been seen in similar
secondaries of cataclysmic variables, which suggests that the high Li
abundance is not due to an anomalous suppression of Li depletion in
close binaries.  SXTs in the quiescent state have
hot advection-dominated accretion flows (ADAFs) in which the ions are
essentially at virial temperature.  At such temperatures, 
Li production via $\alpha-\alpha$ spallation is possible.  We show that
quiescent SXTs can produce sufficient Li via spallation to explain the
observations in V404 Cyg, A0620-00, GS2000+25, and Nova Mus 1991. 
Depending on the Li
depletion time scale in the secondary, which may range between
$10^7-10^9$ yr, the model requires $\sim 10^{-4}-10^{-6}$ of the accreted mass
to be intercepted by the secondary after undergoing Li production and being 
ejected. In the
case of Cen X-4, we can explain the observed Li only if the mass
accretion rate is $\sim 10^{-3}$ times the Eddington rate and if there is
enhanced ejection due to a propeller effect.  We discuss possible
observational tests of this proposal.  Li production during outbursts
could be quite important and may even dominate over the production
during quiescence, but the estimate of the Li yield is uncertain.
We calculate the expected luminosity in
gamma-ray lines due to the production of excited Li and Be nuclei, but
conclude that the line cannot be detected with current instruments.
\end{abstract}


\section{Introduction}

Recently, Martin et al. (1992b, 1994a, 1996) detected Li in the late K-type
secondaries of four soft X-ray transients (SXTs), V404 Cyg, A0620-00,
Cen X-4, and Nova Mus 1991, while
Filippenko et al. (1995) and  Harlaftis et al. (1996) detected Li
in a fifth SXT, GS2000+25.  These detections are
surprising because late K-type stars do not usually have strong Li features
(Brown et al. 1989, Pallavicini et al. 1992).  Lithium is destroyed
in stellar interiors, and the Li in the surface layers of these stars is
expected to be depleted through mixing, diffusion, or post-main
sequence dilution (Martin et al. 1994ab and references therein).  The
high abundances seen in SXTs imply that either there is some Li
production mechanism in these systems or that Li depletion is strongly
suppressed in their secondaries.  Martin et al. (1995) found that
cataclysmic variables (CVs) with late type secondary stars similar to
those in SXTs do not show Li.  This strongly suggests that suppression
of Li depletion in the secondaries is not a viable explanation.  One
must, therefore, take seriously the possibility that the accretion
flows in SXTs produce Li.  Moreover, one must identify a mechanism
which works in the case of accreting black holes and neutron stars,
but not for accreting white dwarfs.

Accretion flows around black holes at low mass accretion rates have
been successfully modeled as hot advection-dominated accretion flows
(ADAFs) in which most of the viscously dissipated energy is retained
within the flow and carried inward rather than being lost through
radiative cooling (Narayan \& Yi 1994,1995b, Abramowicz et al. 1995,
Chen et al. 1995).  The ADAF model has been applied to A0620-00 and
V404 Cyg in their quiescent state (Narayan, McClintock, \& Yi 1996,
Narayan, Barret, \& McClintock 1997).  These studies provide a
convincing explanation of the observed X-ray and optical spectra and constrain
the parameters of the accretion flow, including the mass accretion
rate and the temperature and density of the gas.

In the ADAF model, the ions remain essentially virialized at all
radii, with proton temperatures approaching $\sim 10^{12} K$ near the
central black hole or neutron star (Rees et al. 1982, Narayan \& Yi
1995b).  As a consequence, nuclei with energies $>10$ MeV per nucleon
are abundant and spallation processes become possible (Ramadurai \&
Rees 1985, Jin 1990, Martin et al. 1994ab).  Ramadurai \& Rees (1985)
considered ``ion tori'' around pregalactic massive Pop III
remnants and examined deuterium production and its implication for big
bang nucleosynthesis. Jin (1990) studied the production of $^7$Li and
other light nuclei in ion tori and derived constraints on the light
element enrichment of the Galaxy.  Martin et al. (1992, 1994ab) argued
that the observed high Li abundances seen in SXTs could be explained
by spallation among energetic particles during outbursts of SXTs.

In this paper, we quantitatively examine Li production through
spallation in ADAFs and show that SXTs in quiescence are quite
efficient at producing Li. The Li yield which we calculate is more
than adequate to explain the observed abundances in the black hole
SXTs, V404 Cyg, A0620-00, Nova Mus 1991 and 
GS2000+25, while in the case of the
neutron star SXT, Cen X-4, we need to invoke some assistance from a
propeller mechanism.

\section{Hot Advection Dominated Flows and Lithium Production}
\subsection{$\alpha-\alpha$ Spallation Cross-Section}

The relevant Li production process in hot ADAFs is $\alpha-\alpha$
spallation, $^4$He$(\alpha$,p)$^7$Li, while Li destruction occurs
primarily through the proton initiated process, $^7$Li(p,$\alpha)^4$He
(Rytler 1970, Meneguzzi et al. 1971, Bodansky et al. 1975, Jin 1990). 
The symbols $p$, $\alpha$ refer to H, $^4$He respectively.
For particles with energies much higher than those achieved in ADAFs,
Li production is possible in spallation processes involving heavier
elements such as C,N,O (Meneguzzi et al. 1971, Meneguzzi \& Reeves 1975,
Reeves 1974, Boesgaard \& Steigman 1985, Jin 1990), 
but this channel is not of interest in ADAFs.
The ratio of the production and destruction rates of Li via $\alpha-\alpha$
spallation is given by
$n_{\alpha}n_{\alpha}v_{\alpha\alpha}\sigma_+/n_pn_{Li}v_{pLi}\sigma_-$,
where the $n$'s refer to particle number densities, the $v$'s are
relative particle speeds, and $\sigma_{+(-)}$ are the production
(destruction) cross-sections.  Since we generally have
$n_{\alpha}\gg n_{Li}$, the effect of destruction is negligible
whenever the production cross-section is non-vanishing; destruction
becomes dominant only when production ceases altogether, which
requires the mean particle energy to be less than a few MeV per
nucleon (e.g. Reeves 1974).

The Li production cross-section through $\alpha-\alpha$ spallation is
essentially zero for $E<8.5$ MeV, where $E$ is the mean relative
kinetic energy per nucleon (i.e. the energy per nucleon of one of the particles
as viewed in the rest frame of the other particle).  
Above this energy, the cross-section
increases rapidly, reaching a value $\sim 100$ mb at $E\sim 9$ MeV.
The cross-section decreases again rapidly for $E>15$ MeV, falling to a
few mb at $E\sim 40$ MeV (Meneguzzi et al. 1971, Bodansky et al. 1975,
Jin 1990, and references therein).  For simplicity, we model the
cross-section as
\begin{equation}
\sigma_+(E)\approx 100 (E/10~{\rm MeV})^{-2} ~{\rm mb}, \qquad
E\geq8.5~{\rm MeV}.
\end{equation}
This is the total cross-section for the production of $^7$Li in its
ground state and excited state (at 478 keV) as well as the production
of $^7$Be, via $^4$He$(\alpha$,n)$^7$Be, in its ground state and
excited state (at 431 keV); $^7$Be decays into $^7$Li through electron
capture and is an important channel for Li production
(Meneguzzi et al. 1971, Bodansky et al. 1975).  The cross-sections for
the four species, $^7$Li, $^7$Li$^*$, $^7$Be, and $^7$Be$^*$ (where
the $*$'s represent excited nuclei) are roughly equal (Burcham et
al. 1958, Kozlovsky \& Ramaty 1974, Bodansky et al. 1975 and
references therein).

\subsection{Advection-Dominated Accretion Flows}

The dynamical properties of ADAFs are well understood, and detailed
global solutions as a function of radius, with physically motivated
boundary conditions, have been calculated (Narayan, Kato \& Honma 1997, 
Chen, Abramowicz \& Lasota 1997).  For many purposes, however,
it is sufficient to make use of a simpler self-similar solution
obtained by Narayan \& Yi (1994, 1995b, see also Spruit et al. 1987).
According to this solution, the density, proton temperature, and
radial velocity have the following dependences as a function of the
dimensionless radius $r$ in Schwarzschild units ($r\equiv R/R_S,
~R_S=2GM/c^2, ~M=$ mass of the accreting star),
\begin{equation}
\rho=3.79\times 10^{-5} \alpha^{-1}c_1^{-1}c_3^{-1/2}m^{-1}{\dot m}r^{-3/2},
\end{equation}
\begin{equation}
T=6.66\times 10^{12}\beta c_3 r^{-1} ~{\rm K},
\end{equation}
\begin{equation}
v_R=2.12\times 10^{10} \alpha c_1 r^{-1/2} ~{\rm cm\,s^{-1}}.
\end{equation}
Here $m=M/M_\odot$ is the mass of the star in solar units, ${\dot
m}={\dot M}/{\dot M}_{Edd}=\dot M/1.39\times 10^{18} m ~{\rm
g\,s^{-1}}$ is the mass accretion rate in Eddington units, and
$\alpha$ is the usual viscosity parameter (e.g. Frank et al. 1992);
$\beta$ is the ratio of gas pressure to total pressure, and the
constants $c_1$ and $c_3$ are defined in Narayan \& Yi (1995b).  (Note
that the formula for the ratio of specific heats $\gamma$ given in
equation 2.7 of Narayan \& Yi 1995b should be replaced by
$\gamma=(8-3\beta)/(6-3\beta)$, as shown by Esin 1996).

In the following we retain $\alpha$ as a free parameter, assigning a
value $\alpha=0.3$ whenever we need numerical estimates.  We assume
that $\beta=0.5$, corresponding to gas and magnetic pressure in
equipartition.  For this choice of $\beta$, we have $c_1=1/2$,
$c_3=1/3$.  Assuming that the accreting gas consists of 75\% H and
25\% He by mass, the number densities of H and $^4$He nuclei are given by
\begin{equation}
n_H=5.93\times 10^{19} \alpha^{-1} m^{-1}
\dot m r^{-3/2} ~{\rm cm^{-3}},
\end{equation}
\begin{equation}
n_{\alpha}=4.94\times 10^{18} \alpha^{-1} m^{-1}
\dot m r^{-3/2} ~{\rm cm^{-3}}.
\end{equation}

We assume that the heating rates of different particles in the gas are
proportional to their individual masses, as often assumed for modeling
viscous heating in hot accretion flows (Shapiro, Lightman \& Eardley
1976, Rees et al. 1982).
This assumption (or something similar to it) is critical for the
viability of two-temperature ADAF models.  
Under it, most of the viscous energy 
goes into the ions, and very little goes to the electrons.
If in addition ion-electron coupling via Coulomb collisions is inefficient,
then the electrons decouple from the ions
and cool radiatively to a much lower temperature
than the ions.  This leads to a radiatively inefficient two-temperature ADAF. 

Energy transfer among ions via Coulomb collisions is even more
inefficient than ion-electron energy transfer.  Therefore, if the
various species of ions receive different amounts of energy through
heating, they will not come into thermodynamic equilibrium with one
another.  In fact, it is unlikely that the individual ion species will
achieve a thermal energy distribution among themselves.  Therefore,
when we discuss below the ``temperature'' of ions, we refer merely to
the mean energy of the particles.

If the heating rate is proportional to particle mass as assumed above, 
the mean energy {\it per nucleon} of
the various nuclear species will be the same, namely $3kT/2$. 
We assume this in what follows.
(However, as a practical matter, it makes little difference for this paper
whether different nuclei 
have the same temperature or the same energy per nucleon; we choose the
latter merely because it seems more natural under the assumptions underlying
the ADAF paradigm.)
If we consider two interacting particles in the rest frame of one of the
particles, the mean energy per nucleon of the other particle is $3kT$
and the rms relative speed is $v_r=\sqrt{6kT/m_u}$:
\begin{equation}
E=287r^{-1/2} ~{\rm MeV},
\end{equation}
\begin{equation}
v_r=2.35\times 10^{10} r^{-1/2} ~{\rm cm\,s^{-1}}.
\end{equation}

\subsection{Lithium Production in ADAFs}

For simplicity, we assume here that all pairs of interacting particles
have the same relative energy $E$ and relative velocity $v_r$ as given in
equations (2-7) and (2-8).  Section 3.3 discusses a more detailed
calculation where we use the full particle energy distribution.

The abundance of Li grows as a result of spallation as the accreting
gas flows in.  The change in the abundance over a radial distance
$\Delta R$ is given by
\begin{equation}
{\Delta n_{Li}\over n_{H}}={1\over 2}\sigma_+(E) v_r
{n_{\alpha}^2\over n_H}\Delta t_{flow},
\end{equation}
where $\Delta t_{flow}=\Delta R/v_{R}$.  The factor $1/2$ is to
correct for double counting of $\alpha$ particles.  Since
$\sigma_+\propto E^{-2}\propto T^{-2}\propto r^2$ and $\Delta
t_{flow}= -(R/v_R)\Delta\ln R\sim -r^{3/2}\Delta\ln r$, we have
$\Delta n_{Li}/n_{H}\propto -r^{3/2}\Delta\ln r$.  Thus, most of the
Li is produced at larger radii.  The production switches on suddenly
when $E$ crosses 8.5 MeV at $r_{out}=33.8$, and the rate of production
then decreases as the gas flows in.  This feature means that it is
legitimate to use the self-similar equations (2-2)--(2-4), since the exact
global solutions are very close to the self-similar form at large
radii and show significant deviations only close to the black hole
(Narayan et al. 1997, Chen et al. 1997).  Integrating equation (2-9)
over radius, the total $^7$Li abundance in the accreting gas as it
approaches the black hole is
\begin{equation}
{n_{Li}\over n_H}=\int d\left(n_{Li}\over n_H\right)={1\over 2}
\int_1^{r_{out}} \sigma_+ v_r{n_{\alpha}^2\over n_H} {R_Sdr\over v_R}
=2.13\times10^{-3}{\dot m\over\alpha^2}.
\end{equation}
In terms of mass, the rate of production of Li is
\begin{equation}
\dot M_{Li}=7{n_{Li}\over n_H}0.75\dot M
=1.12\times10^{-2}{\dot m\over\alpha^2}\dot M
=2.47\times10^{-10}{m\dot m^2\over\alpha^2}~M_\odot\,{\rm yr^{-1}},
\end{equation}
where the factor of 7 is for the number of nucleons per $^7$Li
nucleus, and 0.75 is to allow for the fact that only 0.75 of the
accreted mass is in the form of H.

\subsection{Lithium Enrichment of the Secondary}

We assume that a fraction of the accreting mass is ejected
outward in an outflow or wind. This is not unreasonable
as ADAFs have been shown to be
susceptible to outflows/winds (Narayan \& Yi 1994, 1995a), and
there exists direct evidence for ejections in some X-ray binaries
(e.g. Hjellming \& Han 1995, Foster et al. 1996, and references therein).  
We further assume that a fraction
of the outflowing material is intercepted by the secondary.
Thus, we write the fraction of the accreting mass that reaches the
secondary as $F_{esc}\Omega$, where $\Omega$ is the solid angle of
the secondary as viewed from the accreting star.  We treat
$F_{esc}$ as a parameter, and note that there is considerable
uncertainty in its value.

There is at present no reliable physical decription of
outflows/winds from accreting black holes.  Therefore, the total
fraction of the accreting mass which flows out is not known.
Further, the angular distribution of the outgoing mass is uncertain
and it is not clear how much of this mass flows in the direction of
the secondary.  Finally, the capture probability on the secondary
is also uncertain since it could be modified by a stellar wind
or a stellar magnetosphere.  We take the
point of view that any value of $F_{esc}\ll1$ is ``reasonable,''
while a value of $F_{esc}\to1$ is too optimistic (except in the
propeller case considered in \S3.3).

The rate at
which Li is deposited on the surface of the secondary is given by
\begin{equation}
\dot M_{Li,+}=F_{esc}\Omega\dot M_{Li}
=2.47\times10^{-12}{m\dot m^2\over\alpha^2}F_{esc}\Omega_{-2}
~M_\odot{\rm yr^{-1}},
\end{equation}
where $\Omega_{-2}=\Omega/10^{-2}$.  In writing this result we assume
that most of the outflow occurs from small radii, inside the radius
$\sim30$ where the bulk of the Li synthesis takes place.

The Li deposited on the secondary is depleted by destruction
processes in the star.  The depletion time scale is
somewhat poorly determined for stars of various kinds (e.g. Boesgaard \&
Steigman 1985).  The original Pop I Li abundance of 
$\sim 10^{-9}$ (Boesgaard \& Steigman 1985, Reeves et al. 1990) 
with which a star begins its life
decreases during several stages of stellar evolution. (i)
During the pre-main sequence phase, vigorous convective transport
could substantially deplete the surface Li.  In young stellar
clusters, the depletion is observed to depend on stellar type and
there is a significant spread of the abundance from very low values
all the way to the primordial value.  The depletion time scale appears
to be $\sim10^7-10^8$ yr (Martin et al. 1992ab and references
therein).  (ii) During the main sequence phase of K dwarfs, the
depletion time scale appears to be as short as a few $\times 10^8$ yr
(e.g. Boesgaard \& Steigman 1985, Thorburn et al.  1994, Garcia-Lopez
et al. 1994), as suggested by low observed Li abundances (Brown 1989,
Pallavicini et al. 1992, Martin et al. 1994b).  However, the main
sequence depletion time scale for F and G type dwarfs and subgiants
may be as long as $\sim 10^9$yr (e.g. Duncan 1981).  (iii) G and K
giants appear to deplete Li by a large factor $\sim 10^3$ on a time
scale $\sim 5\times 10^7$ yr (Pilachowski et al. 1984, Boesgaard \&
Steigman 1985). The depletion time scale of evolved stars such as
stripped giants in V404 Cyg and Cen X-4 is poorly known. 

We take the depletion time scale to be another
free parameter, with a value in the range $10^7-10^9$ yr, and we
scale our results to a fiducial time scale of $10^8$ yr.  If the
depletion time scale is as long as $\sim 10^9$yr as suggested by
the work of Pinsonneault et al. (1992), then 
our estimates may be considered conservative.

Let $(n_{Li}/n_H)_{-9}$ be the Li abundance in the envelope of the
secondary in units of $10^{-9}$.  The rate of destruction of Li is
given by
\begin{equation}
\dot M_{Li,-}=7\times10^{-9}\left({n_{Li}\over n_H}\right)_{-9}{0.75M_{env}
\over t_D} =5.25\times10^{-18}{M_{env,-1}\over t_{D8}}
\left({n_{Li}\over n_H}\right)_{-9} ~M_\odot\,{\rm yr^{-1}},
\end{equation}
where $M_{env,-1}$ is the mass of the secondary's envelope in units of
$0.1M_\odot$, and $t_{D8}$ is the depletion time in units of $10^8$yr. 
It is likely that in some of our systems the
envelope mass is much lower than $0.1M_\odot$
(cf. Pinsonneault et al. 1992).  We thus err again on the
side of being conservative in our choice of scaling for $M_{env}$.

If the Li abundance in the secondary has reached a steady state,
the enrichment and depletion rates should be equal.  Equating (2-12)
and (2-13), we then obtain the escape fraction $F_{esc}$ needed in
order to explain the observed Li abundance,
\begin{equation}
F_{esc}=2.13\times10^{-6}{\alpha^2\over m\dot m^2}
{M_{env,-1}\over \Omega_{-2}t_{D8}}\left({n_{Li}\over n_H}\right)_{-9}.
\end{equation}

If the total duration of the accretion flow is shorter than the
depletion time of the secondary, then we can neglect Li destruction
and assume that the secondary retains all the Li deposited on it
during the life of the system as an X-ray binary.  In this case,
equation (2-14) is replaced by
\begin{equation}
F_{esc}=4.71\times10^{-5}{\alpha^2\over \dot m}
{M_{env,-1}\over \Omega_{-2}\Delta M_{-1}}\left({n_{Li}\over n_H}\right)_{-9},
\end{equation}
where $\Delta M_{-1}$ is the total mass transferred from the secondary
to the primary in units of $0.1M_\odot$.  (We have made use of the
relation $\dot M = 2.21\times10^{-8}m\dot m ~M_\odot\,{\rm yr^{-1}}$.)

\subsection{Gamma-Ray Line Emission}
Roughly half the $^7$Li and $^7$Be nuclei produced via $\alpha-\alpha$
spallation are in an excited state.  When these nuclei make a
transition to the ground state they emit gamma-rays at $478$ keV
($^7$Li$^*$) and $431$ keV ($^7$Be$^*$) respectively (Kozlovsky \&
Ramaty 1974).  If this line emission could be detected it would
provide strong support for the spallation scenario (Martin et
al. 1992b, 1994b).  Using the Li production rate estimated in
section 2.3, we calculate the luminosity in gamma-ray lines to be
\begin{equation}
L_{\gamma}\sim 4.86\times 10^{32} {m\dot m^2\over\alpha^2}
~{\rm erg\,s^{-1}},
\end{equation}
where we have assumed that one quarter of the produced nuclei emit 478
keV photons and one quarter emit 431 keV photons.  This is a very low
gamma-ray luminosity, especially considering the fact that the ADAF
solution is valid only for low mass accretion rates, $\dot
m<(0.3-1)\alpha^2$ (Narayan \& Yi 1995b).  Setting $\dot m=\alpha^2$
and taking $\alpha=0.3$, we obtain a maximum gamma-ray luminosity of
\begin{equation}
L_{\gamma,max}=4.37\times10^{31}m ~{\rm erg\,s^{-1}}.
\end{equation}
Even with $m\sim20$, the maximum likely mass of a black hole in an
X-ray binary, the luminosity is too low to be detected with current
detectors.  Furthermore, the time scale for gamma-ray emission from
excited nuclei, of the order of days to weeks, is fairly long 
(mainly determined by the electron capture time scale for $^7$Be nuclei;
cf. Browne \& Firestone 1986) and so most of the nuclei
are likely to disappear into the black hole before they can decay and
emit gamma-rays. (This is not an issue for neutron star SXTs.)
The line may possibly be within the limits of the SPI Ge
spectrometer on the INTEGRAL mission.

When ${\dot m}>\alpha^2$, the accretion flow is likely to be in the
form of a cool geometrically thin accretion disk (e.g. Frank et
al. 1992).  In such disks, spallation can occur (if at all) only in
non-thermal flares (e.g. Field \& Rogers 1993).  If flare activity is
large enough and if it produces high energy alpha particles ($>10$ MeV
per nucleon) in a dense environment, then in principle one might have
a detectable flux of gamma-ray lines.  But such a model does not fall
within the scope of the ADAF paradigm considered here.

\section{Application to Soft X-Ray Transient Systems}

In this section we apply the above estimates to the SXT systems
with high observed Li abundances.  The solid angles of the secondaries
are given by $\Omega=\pi R_{sec}^2/4\pi a^2$, where $R_{sec}$ is the radius of
the secondary and $a$ is the separation of the two stars.  We estimate
$R_{sec}/a$ using the fitting formula of Eggleton (1983),
\begin{equation}
{R_{sec}\over a}={0.49q^{2/3}\over 0.6q^{2/3}+\ln(1+q^{1/3})},
\end{equation}
where $q=M_{sec}/M$ is the mass ratio between the secondary and
primary.

\subsection{V404 Cyg}
The dynamical parameters of this black hole SXT are relatively well
constrained: $M\sim12M_{\sun}$, $M_{sec}\sim0.7M_\odot$
(Shahbaz et al. 1994), $q=0.0583$,
which give $R_{sec}/a=0.176$, $\Omega_{-2}=0.777$ .
The observed Li abundance in the secondary is
$\log(n_{Li}/n_H) =-9.4$ (Martin et al. 1994a).  Narayan, Barret \&
McClintock (1997)
fitted the X-ray and optical spectrum of V404 Cyg in quiescence and
estimated a mass accretion rate of $\dot m=0.0046$ for $\alpha=0.3$.
If we take the mass in the envelope of the secondary to be
$0.2M_{sec}$, and assume that the Li in the secondary is in steady
state, then equation (2-14) gives
\begin{equation}
F_{esc}\sim 5.4\times 10^{-4} \left(M_{env}\over 0.14M_\odot\right)
\left(10^8{\rm yr}\over t_D\right).
\end{equation}
Alternatively, if we assume that the depletion time is longer than
the X-ray lifetime of the system, we obtain from equation (2-15)
\begin{equation}
F_{esc}\sim 4.7\times 10^{-4}
\left(M_{env}\over 0.14M_\odot\right)
\left({0.14M_\odot\over\Delta M}\right).
\end{equation} 
With either estimate we see that there needs to be only a very small
level of mass ejection, of the order of $0.1\%$ of the mass accretion
rate, in order to contaminate the secondary with the observed level of
Li.  Even if the Li depletion time in the secondary is as short as
$10^7$ yr, the fraction of escaping material still has to be only
about 1\% of the accreted mass.  In fact, since Pinsonneault et
al. (1992) suggest a long depletion time $\sim10^9$ yr, the parameter
$F_{esc}$ may be as small as $10^{-4}$.  Thus, spallation in the hot
ADAF during the quiescent state of V404 Cyg is a very promising
mechanism to explain the observed Li excess in the secondary.

\subsection{A0620-00 and Other Similar Systems} 

We adopt the following system parameters: $M=6M_{\sun}$, $M_{sec} =0.5
M_\odot$ (Barret, McClintock \& Grindlay 1996), $q=0.0833$, which give
$R_{sec}/a=0.196$, $\Omega_{-2}=0.960$.  The observed Li abundance is
$\log(n_{Li}/n_H) =-10$ (Martin et al. 1994a).

Narayan, McClintock \& Yi (1996) 
fitted the X-ray and optical spectra of A0620-00
and estimated $\dot m = 2\times10^{-4}$ for $\alpha=0.3$.
The steady state value of $F_{esc}$ is then
\begin{equation}
F_{esc}\sim 0.083 \left(M_{env}\over 0.1M_\odot\right)
\left(10^8{\rm yr}\over t_D\right).
\end{equation}
Note, however, that the Narayan et al. (1996) model was based on a
black hole mass of $4.4M_\odot$ and corresponded to $\beta=0.95$.  A
reanalysis, with $M=6.1M_\odot$, $\beta=0.5$, and making use of the
improved modeling techniques described in Narayan et al. (1997), gives
$\dot m=9.7\times 10^{-4}$.  
For this value of $\dot m$, assuming steady state,
we find
\begin{equation}
F_{esc}\sim 3.5\times10^{-3} \left(M_{env}\over 0.1M_\odot\right)
\left(10^8{\rm yr}\over t_D\right),
\end{equation}
while for the case when depletion can be neglected we find
\begin{equation}
F_{esc}\sim 4.6\times 10^{-4}
\left(M_{env}\over 0.1M_\odot\right)
\left({0.1M_\odot\over\Delta M}\right).
\end{equation}
As in the case of V404 Cyg, we see that we need about 0.1\% of the
accreted mass to be ejected (1\% if $t_D=10^7$ yr) in order to produce
the observed level of Li in the secondary.

The black hole SXTs, GS2000+25 and Nova Mus 1991, are fairly similar
to A0620-00 in their binary parameters: $M=6-14M_\odot$,
$M_{sec}=0.2-0.6M_\odot$ (Filippenko et al. 1995, Harlaftis et
al. 1996, Barret et al. 1996).  The observed Li abundances in the
secondaries also agree to within an order of magnitude (Harlaftis et
al. 1996).  There are no reliable models yet of these systems in
quiescence, and we do not have an independent estimate of $\dot m$.
The observed Li requires that $\dot m$ in quiescence should be similar
to the values we have estimated for V404 Cyg and A0620-00.

\subsection{Cen X-4}
This neutron star SXT has the highest observed Li abundance among all
SXTs, $\log(n_{Li}/n_H)=-8.7$ (Martin et al. 1994a).  We take the
following system parameters: $M=1.4M_{\odot}$, $M_{sec}=0.1M_\odot$
(McClintock \& Remillard 1990), $q=0.0714$, which give
$R_{sec}/a=0.187$, $\Omega_{-2}=0.874$.

The mass accretion rate in the system is uncertain.  The observed
quiescent X-ray luminosity of $\sim 2.4\times 10^{32} ~{\rm
erg\,s^{-1}}$ (Asai et al. 1996) implies a very low $\dot
m\sim10^{-6}$.  Equations (2-14) and (2-15) show that it is impossible
with such a low $\dot m$ to produce the observed level of Li.  Could
$\dot m$ be substantially larger?

Since Cen X-4 is a neutron star system, we need to consider the
possibility that the star may have a moderately strong surface
magnetic field.  Coherent pulsation with a period of $P_*=31.28$ ms
may have been observed in this system in quiescence (Mitsuda et al. 1996).
The signal is most likely due to the rotation of the neutron
star, and indicates the likely presence of a magnetosphere (e.g. Frank
et al. 1992). 
ADAFs have been shown to have a substantially sub-Keplerian rotation 
(Narayan \& Yi 1994, 1995b), 
$\Omega=c_2(GM/R^3)^{1/2}$, where the coefficient $c_2$ is given in 
Narayan \& Yi (1995b). For $\alpha=0.3$, $\beta=0.5$, we obtain $c_2=0.417$.
Taking the measured spin period in Cen X-4, 
the corotation radius is 
\begin{equation}
r_c=R_c/R_s=c_2^{2/3}(GMP_*^2/4\pi^2)^{1/3}/(2GM/c^2)\sim 22,
\end{equation}
while the magnetospheric radius (or Alfven radius)
for a surface magnetic field strength 
$B_*$ is
\begin{equation}
r_A=R_A/R_s\sim 44\left(B_*\over 10^9G\right)^{4/7}\left({\dot M}\over
10^{15} g/s\right)^{-2/7}.
\end{equation}
If $B_{*}\sim 10^9({\dot M} /10^{15} ~{\rm g\,s^{-1}})^{1/2}G$ (a
reasonable value based on the field strengths seen in millisecond
pulsars), then it is quite possible that Cen X-4 in quiescence has its
magnetospheric radius somewhat outside the corotation radius.  
If $r_A$ is sufficiently large ($>50$), the
system could be in the ``propeller regime'' (Illarionov \&
Sunyaev 1975), where the bulk of the accreted material is stopped by
the magnetic field and flung out by centrifugal action (Asai et
al. 1996, Tanaka \& Shibazaki 1996).  

The existence of a propeller
enhances the predicted Li in the secondary in two ways.  First, if
there is a propeller, the mass accretion rate is much higher than that
inferred from the X-ray luminosity, since only a very small fraction
of the accreting material actually reaches the neutron star.  This
obviously increases the Li yield in the accretion flow (eq 2-11).
Second, the propeller action ensures that essentially all the
accreting material is thrown out, so that we expect the parameter
$F_{esc}$ to be essentially of order unity.  Thus, a much
larger fraction of the Li produced in the accretion ends up on the
secondary.  There is a counter-effect, however.  If the magnetospheric
radius is larger than the critical radius $r_{out}=33.8$ calculated in
sec. 2.3, then very few alpha particles achieve the energy ($\sim8.5$
MeV per nucleon) needed for spallation, and the Li yield is less than
in the black hole case.

Let us assume that the bulk of the mass flow is stopped by the magnetospheric
pressure at $r=r_A$ and expelled through the propeller effect.  
We take $r_A\sim 50$, the radius at which the centrifugal action is just able
to drive the accreted
material to infinity. Although for $r\ge 33.8$ the mean energy per nucleon is 
below the $\alpha-\alpha$
spallation threshold, some Li can still be produced by alpha particles
in the high energy tail of the particle energy distribution.  To
estimate the reaction rate we need to do a more careful calculation
than we did in sec. 2.3.  Adopting a Maxwellian distribution (this is
just a convenient model and the real distribution may be quite different), the
effective interaction rate is given by
\begin{equation}
<\sigma_+ v>=\int n(E)\sigma_+(E)v(E)dE,
\end{equation}
where
\begin{equation}
n(E)dE={2\over \sqrt{\pi}(kT)^{3/2}}\exp(-E/kT)E^{1/2}dE
\end{equation}
(e.g. Cox \& Giuli 1968), and $v(E)=\sqrt{2m_uE}$.  Carrying out the
integral over energy and radius, we find that the abundance of Li when
the accretion flow reaches $r=50$ is
\begin{equation}
{n_{Li}\over n_H}=2.33\times 10^{-5} {{\dot m}\over\alpha^2}.
\end{equation}
The Li yield is lower by a factor $\sim 90$ than for the case considered
in sec. 2.3 (eq. 2-10).  The inefficiency arises because the flow is
truncated before it can reach the optimum radius ($r\sim30$) for
spallation.  Incidentally, if we allow the flow to extend down to
$r=1$, the present more detailed calculation gives a coefficient of
$2.66\times10^{-3}$ in equation 2-10, instead of $2.12\times10^{-3}$;
thus, the simplifying assumption made in sec. 2.3 leads to an error of
about 20\%.  
If we truncate the accretion flow at a radius $r\sim10^3$, as appropriate
for an accreting white dwarf, there is no Li production at all by
spallation.  Thus, the absence of Li in CVs (Martin et al. 1994a) is
naturally explained in this model.

With the lower Li yield given in equation (3-11), equation (2-14) is
modified to
\begin{equation}
F_{esc}=2.00\times10^{-4}{\alpha^2\over m\dot m^2}
{M_{env,-1}\over \Omega_{-2}t_{D8}}\left({n_{Li}\over n_H}\right)_{-9}.
\end{equation}
Let us assume that $F_{esc}=1$ in Cen X-4 because of the propeller
effect.  Then, setting $\alpha=0.3$ and substituting the values of the
various other quantities, we can solve for $\dot m$:
\begin{equation}
\dot m = 2.3\times10^{-3}\left({M_{env}\over0.02M_\odot}\right)^{1/2}
\left({10^8{\rm yr}\over t_D}\right)^{1/2}.
\end{equation}
Alternatively, if we assume that the depletion time is very long and
use the equivalent of equation (2-15), we obtain
\begin{equation}
\dot m = 8.1\times10^{-4}\left({M_{env}\over0.02M_\odot}\right)
\left({0.02M_\odot\over \Delta M}\right).
\end{equation}
The two estimates are roughly consistent with each other, and in fact
give quite a reasonable value of $\dot m$, since it is quite similar
to the values of $\dot m$ we have estimated in V404 Cyg and A0620-00.
Black hole SXTs and neutron star SXTs are quite similar to each other
in many respects.  We might, therefore, expect their mass accretion
rates (scaled to the Eddington value) to be similar, both in
quiescence and outburst.  As supporting evidence we note that Cen X-4
and A0620-00 seem to be similar to each other in their outbursts.  Cen
X-4 has had two outbursts in the last 30 years with a total X-ray
output of about ${\rm few} \times10^{44}$ ergs, which is fairly
similar to the energy output of A0620-00 during a similar period
(Tanaka \& Shibazaki 1996).  This suggests that the mass storage rates
in the two systems are comparable.  It is reasonable to think that the
quiescent accretion rate in Cen X-4 also is roughly the same as in
A0620-00, i.e. $\dot m\sim10^{-3}$.  Since this is more-or-less the
value we need to explain the observed Li in Cen X-4, we argue that the
scenario is consistent.

\subsection{Soft X-Ray Transients in Outbursts}

In addition to the quiescent state which we have considered so far, Li
may also be produced during periods of more rapid mass accretion in
outbursts (see Tanaka \& Shibazaki 1996 for a discussion of SXT outbursts). 
The hot ADAF
solution on which we have based our estimates exists only for ${\dot
m}<\dot m_{crit}\sim(0.3-1)\alpha^2$ (Narayan \& Yi 1995b).  We need,
therefore, to determine exactly when during an outburst the accretion
is in the form of an ADAF.  At the peak of the outburst, the mass
accretion rate in SXTs approaches the Eddington limit, $\dot m\to1$.
During this period the accretion will most likely be in the form of a
thin disk (cf. Narayan 1996) and therefore not suitable for producing
Li. However, both when the system is on its way up to the peak and on
its way down from the peak, the flow will go through a period of
advection-dominated accretion with $\dot m$ close to the limiting
$\dot m_{crit}$.  The rise to outburst is usually quite rapid and not
very interesting, but the decline is often slower, and it is likely
that SXTs linger around $\dot m\sim\dot m_{crit}$ for a reasonable
period of time during decline.  During this period, Li synthesis could
be particularly efficient (since eq. 2-11 shows that Li production
varies as $\dot m^2$.).  Using the subscript ``high'' to refer to
episodes of $\dot m\sim\dot m_{crit}$ and the subscript ``low'' for
the quiescent state, we estimate the relative Li production in the two
phases to be
\begin{equation}
{M_{Li}(high)\over M_{Li}(low)}\sim 
\left(\Delta t_{high}\over \Delta t_{low}\right)
\left({\dot m}_{high}\over {\dot m}_{low}\right)^2
\left(\alpha_{high}\over \alpha_{low}\right)^{-2}
\left(F_{esc,high}\over F_{esc,low}\right),
\end{equation}
where $\Delta t$, $\alpha$, and $F_{esc}$ refer to the duration, the
viscosity parameter, and the ejection fraction.  Taking typical time
scales, $\Delta t_{high}\sim 0.1$ yr and $\Delta t_{low}\sim 30$ yr,
and assuming that $\alpha$ and $F_{esc}$ in the two phases are the
same, we find that the amount of Li produced in the high phase
exceeds that produced in the low phase if
\begin{equation}
\left({\dot m}_{high}\over {\dot m}_{low}\right)>
17\left(\Delta t_{low}/\Delta t_{high}\over300\right)^{1/2}.
\end{equation}

In view of the values of $\dot m_{low}$ we have estimated (see
secs. 3.1--3.3) and the likely value of $\dot m_{high}$ ($\sim
0.03-0.1$ for $\alpha=0.3$), we infer that Li production during
outbursts could be competitive with production during quiescence, and
might even dominate.  To see this another way, we follow the methods
described in sec. 2.4 and estimate the equilibrium abundance of Li in
the secondary purely as a result of outbursts:
\begin{equation}
{n_{Li}\over n_H}\sim {\Delta M_{Li}t_D\over 7\times 0.75 M_{env}t_{rec}}
\sim 1.4\times 10^{-10} 
m\left(\Omega\over 10^{-2}\right)
\left( F_{esc}\over10^{-3}\right)
\left( t_{D}\over10^8~{\rm yr}\right)
\left(0.1M_\odot\over M_{env}\right)
\left(300\over \Delta t_{low}/\Delta t_{high}\right),
\end{equation}
where we have used $\alpha=0.3$ and ${\dot m}_{high}=\alpha^2=0.09$.
We find that the predicted Li abundance due to outbursts is comparable
to the observed values.

Thus, we conclude that Li production during outbursts could be
important and has to be considered seriously.  However, the exact
variation of $\dot m$ during the decline from outburst is not well
understood and it is not clear exactly at which stage of the decline
the accretion switches from a thin disk to an ADAF.  In view of this
uncertainty, the estimates given here are less reliable than the
values given earlier for quiescent SXTs.

\subsection{Observing Gamma-Ray Lines from Excited Li}

In some black hole X-ray binaries such as Nova Mus 91 and 1E
1740.7-2942, a gamma-ray line feature near $\sim 480$ keV has been
reported (e.g. Goldwurm et al. 1992, Bouchet et al. 1991), which is
interestingly close to the gamma-ray emission line (478 keV) expected
from excited $^7$Li in spallation (Martin et al. 1992b, 1994ab).  As
we have shown in sec. 2.5, however, gamma-ray line emission from ADAFs
in X-ray binaries has a maximum luminosity of only $\sim10^{33}~{\rm
erg\,s^{-1}}$, even if we ignore the loss of Li into the black hole,
whereas the line detected in Nova Mus 91 had a luminosity of $10^{37}
~{\rm erg\,s^{-1}}$ for an assumed distance of 5 kpc.  Another problem
is that the line in Nova Mus 91 was observed at a time when the system
was either in the ``high'' or ``very high'' state.  These states are
likely to involve accretion via a thin accretion disk (cf. Narayan
1996) for which the present analysis is not relevant.

\subsection{Lithium Production in Other Black Hole Systems}

Cyg X-1 is a bright X-ray binary which very likely undergoes accretion
via an ADAF (at least in the ``low state'', cf. Narayan 1996).  The
X-ray luminosity of the source is $\sim{\rm few}\times 10^{37} ~{\rm
erg\,s^{-1}}$, which corresponds to $\dot m\sim 0.1$ for a black hole
mass of $\sim10M_\odot$.  By the estimates given in this paper, Cyg
X-1 must be producing a large quantity of Li, of which substantial
amounts must be intercepted by the secondary.  If the depletion time
is not different from that in other stars, and if the Li is not swept
away by the strong wind from the star, the abundance of Li in the
secondary must be fairly high.  Unfortunately, the star is too hot to
reveal Li in its spectrum, and so this prediction cannot be tested.

At the current level of activity, ${\dot m}\sim 10^{-3}$ (Narayan, Yi
\& Mahadevan 1995), the Galactic center source Sgr A$^*$ ($M\sim
10^6M_{\sun}$) would produce $\sim 30 M_{\sun}$ of $^7$Li over its
life time of $\sim 10^{10}$ yr (cf. eq 2-10), of which about
$0.03M_\odot$ would be ejected into the surrounding ISM, assuming
$F_{esc}\sim10^{-3}$.  If the source was more active in the past than
it is at present, the Li enrichment near the Galactic center could be
even larger (since $\dot M_{Li}$ varies quadratically with $\dot m$,
see eq 2-11).  If the ejected Li does not diffuse too far from the
Galactic Center there may be some evidence for excess Li in newly
formed stars in the region.  At the current level of activity, the
gamma-ray line emission from Sgr A$^*$ is expected to be only
$L_{\gamma}\sim 5\times 10^{33}~{\rm erg\,s^{-1}}$.

\section{Discussion}

In this paper, we have considered Li production through spallation in
hot ADAFs in SXTs, concentrating primarily on the quiescent state of
SXTs (secs. 2.1--2.4).  The ADAF paradigm has been applied
successfully to two black hole SXTs, V404 Cyg and A0620-00, and
provides a good description of the observed spectra in these systems
in quiescence (Narayan et al. 1996, 1997).  The models are well
constrained by the observations and provide all the parameters
necessary in these two SXTs for a quantitative estimate of the Li
yield via spallation.  The only uncertainty at this point concerns the
parameter $F_{esc}$, which is defined such that $F_{esc}\Omega$ is the
fraction of the accreting material which reaches the secondary via an
outflow, where $\Omega$ is the solid angle subtended by the secondary
as viewed from the accreting star.  We find that we can fit the
observed Li abundances in the secondaries of V404 Cyg and A0620-00 if
we assume that $F_{esc} \sim 10^{-3}(10^8 ~{\rm yr}/t_D)$, where $t_D$
is the time scale on which Li is depleted in the envelope of the
secondary.  Since $t_D$ is expected to be in the range $10^7-10^9$ yr,
with the longer time scale more likely (Pinsonneault 1992), the
required ejection fraction is small, making the scenario quite
plausible.  Moreover, we find that we need similar values of $F_{esc}$
in V404 Cyg and A0620-00, which shows that the model is consistent.

A natural consequence of our model is that Li synthesis takes place
only in accretion flows around black holes and neutron stars, but not
around white dwarfs.  The spallation reactions need about 10 MeV per
nucleon, and white dwarf accretion flows do not reach such
temperatures even when they are advection-dominated.  Martin et
al. (1995) found that CVs with late type secondary stars similar to
those in SXTs do not show Li (Martin et al. 1994a).  This is
consistent with our model.

We find that Li production during outbursts of SXTs could be quite
important, and may perhaps even dominate over the production during
quiescence (sec. 3.4).  However, it is not known exactly when during
outburst the accretion occurs as an ADAF and when as a thin disk.
Since the high temperatures needed for spallation are present only in
ADAFs, the estimate of the Li yield is somewhat uncertain.

In principle, Li could be produced even when the accretion is via a
thin disk (say at the peak of an outburst) in nonthermal flares (Field
\& Rogers 1993) or in an active corona.  Another possibility is that
the bulk motion of the ejected material (either during quiescence or
outburst) may be fast enough that when the ejected alpha particles
reach the secondary they produce spallation reactions in situ in the
envelope of the secondary (Rytler 1970).  The uncertainties in such
models are, however, quite severe, and it is very hard to make
quantitative estimates of the Li yield.  Yet another possibility is
that supernova explosions of the progenitor stars might contaminate
the secondaries with Li (Dearborn et al. 1989).  However, in some models the
progenitors of black holes collapse 
without explosions (e.g. Woosley 1993 and references therein).
We do not expect Li contamination of the secondary in such models.

While our spallation scenario works well for quiescent black hole
SXTs, a direct application of the model to the neutron star SXT, Cen
X-4, fails by many orders of magnitude.  The X-ray luminosity of Cen
X-4 in quiescence is very low.  As a result, when we determine the
mass accretion rate $\dot m$ directly from the luminosity, we find
that the estimated Li production is extremely low.  One solution could
be that Cen X-4 produces most of its Li during outbursts.  We,
however, prefer a second solution, namely that Cen X-4 may be
accreting via a propeller mode (Illarionov \& Sunyaev 1975), as argued
by Asai et al. (1996) and Tanaka \& Shibazaki (1996).

We suggest that Cen X-4 has quite a high $\dot m\sim0.002$ in
quiescence, similar to the accretion rate we have estimated in V404
Cyg and A0620-00 from spectral fitting (secs. 3.1, 3.2).  However,
most of the accreting mass is flung out by the centrifugal action of
the rotating neutron star. Since only a tiny fraction of the mass
accretes on the neutron star, the low X-ray luminosity is explained.  
Furthermore, since nearly all the accreting mass flows
out, the amount of Li reaching the secondary is larger than in the
black hole systems.  Thus, the propeller model naturally explains the
unusually high abundance of Li in Cen X-4.

One could make a plausible argument that quiescent neutron star SXTs
are especially likely to be in the propeller regime.  A typical SXT
has its mass accretion rate varying by three or more orders of magnitude
between quiescence and outburst.  By equation (3-8), the
magnetospheric radius $r_A$ should vary by nearly an order of
magnitude, moving in during outburst and moving out in quiescence.
Let us assume that the neutron star attains some kind of equilibrium
spin period appropriate to its mean mass accretion rate.  During
outburst, we will have $r_A<r_c$ (the corotation radius,
cf. eq. 3-7) and the neutron star will undergo spin-up.  However,
during quiescence, we expect $r_A>r_c$ and it is quite likely that we
will have spindown via propeller action.

In our model, Cen X-4 in quiescence has a large mass outflow rate,
$\dot M_{out}\sim2\times10^{15}~{\rm g\,s^{-1}}$, and the outflow
velocity is on the order of the rotation speed at the magnetospheric
radius, $v_{out}\sim0.1c$.  These estimates are based on the tentative
rotation period of the neutron star, $P_*=31.28$ ms, 
identified by Mitsuda et al. (1996).
The presence of an energetic outflow in Cen X-4 may
perhaps be detectable.  For instance, the gas may produce weak radio
emission via synchrotron radiation of shock-accelerated electrons.
Cen X-4 was detected as a $\sim10$ mJy radio source during one of its 
outbursts (Hjellming et al. 1988). Conceivably, the source may be visible
even in quiescence as a much weaker radio source.  Another possibility
is that the ejected material may produce nebular emission when it
shocks with the ISM.  We note in this connection that Kulkarni \&
Hester (1988)  detected an H$\alpha$ nebula around the radio pulsar,
PSR 1957+20; the nebula in that case arises from the interaction of the pulsar
wind with the ISM.  The kinetic energy flux in PSR 1957+20 is
estimated to be about $10^{35}~{\rm erg\,s^{-1}}$, while in Cen X-4 we
estimate a flux of $\sim10^{34}~{\rm erg\,s^{-1}}$.  Because of the
propeller effect, we expect Cen X-4 in general to have stronger evidence of
outflow-related activity than A0620-00, even though the two systems
have similar $\dot m$. (In fact, in physical units, A0620-00 has a higher 
${\dot M}$ than Cen X-4 because of its larger mass.)

Another consequence of our model of Cen X-4 is that the X-ray pulsations
detected in this system (Mitsuda et al. 1996)
should reveal a secular spin-down of the neutron
star. Using our model parameters and a neutron star moment of inertia 
$I_*\sim 10^{45}~{\rm  g\,cm^2}$, we estimate the spindown time scale to be 
$P_*/\dot P_*\sim I_*\Omega_*/{\dot M}\Omega(R_A)R_A^2\sim 10^8$ yr.

One interesting point is that Cen X-4, which has the highest Li
abundance among the three SXTs studied so far (and indeed one of the
highest abundances seen in any star), may be relatively inefficient at
producing Li.  According to our model, the magnetospheric radius in
this system is fairly large, $r_A\sim50$, and lies outside the optimal
radius $\sim30$ for Li spallation.  Thus, for the given $\dot m$, the
accretion flow in Cen X-4 produces about 90 times less Li than an
equivalent black hole system would (sec. 3.3, but note that the argument
is based on the tentative rotation period of 31.28 ms measured by
Mitsuda et al. 1996).  If we could find
another neutron star SXT, with a weaker magnetic field than Cen X-4
such that $r_A<30$, then the Li yield would be substantially higher.
If the system were to have an active propeller in quiescence (which we
argued earlier is likely), then we could easily imagine a steady state Li
abundance in the secondary on the order of $n_{Li}/n_H\sim10^{-8}-10^{-7}$.  
The discovery of such an object would prove
beyond any reasonable doubt that the Li in SXT secondaries is produced
by the accretion flow, and would rule out the alternative hypothesis
that the Li is a fossil left-over from the initial material of the
star.  If a neutron star in an SXT has a stronger magnetic field than
Cen X-4 and if $r_A$ exceeds $\sim60$ say, then the Li yield would be
vanishingly small.  Thus, we expect an inverse correlation between the
magnetic field strength (or equivalently the equilibrium spin period
of the neutron star) and the Li abundance of the secondary.  The
correlation is not likely to be linear, however, but more in the
nature of a step function.

Two outstanding uncertainties in the spallation scenario we have
outlined should be mentioned.

(i) The Li depletion time scale in the secondary is poorly
constrained.  While we have shown that the model works even with a
time scale as short as $\sim 10^7$ yr, as is commonly assumed for K
type main sequence stars, it would be useful to have a better handle
on this parameter. 
For instance, if the depletion time scale is much
longer than the age of the binary system (Duncan 1981, Pilachowski et
al 1984, Boesgarrd \& Steigman 1985, Pinsonneault et al. 1992), 
say as a result of tidal effects (cf. Zahn \& Bouchet 1989, Zahn 1994), 
then it is conceivable that the Li we see
in SXTs is just the Li with which the secondary was originally formed.
This hypothesis is not very attractive in view of the fact that
secondaries in CVs do not have detectable levels of Li (Martin et
al. 1995), but cannot be ruled out conclusively at present.

(ii) A major uncertainty in our model is the angular distribution
of the ejected material.  If most of the ejection occurs along the
poles, then even the small values of $F_{esc}$ which we require in our
scenario may be difficult to achieve.  A further uncertainty is that
ejecta from the accretion flow may be unable to penetrate the
secondary star if the latter has a magnetosphere or an active wind.
One positive feature of the model is that outflows are considered
natural and even likely in ADAFs (Narayan \& Yi 1994, 1995a).
In view of these uncertainties, our results can be stated that we
need a fraction $\sim 10^{-5}(10^8$yr$/t_D)$ of the accreting mass
be intercepted by the secondary stars in V404 Cyg and A0620-00 after
Li production has taken place. In the case of Cen X-4 we need 
$\sim 10^{-2}(10^8$yr$/t_D)$ to reach the secondary. 

Martin et al. (1992b, 1994b) made the interesting suggestion that
gamma-ray lines from excited Li and Be nuclei produced by spallation
may be detectable and may in fact be the origin of a spectral line at
480 keV detected in Nova Mus 91 and 1E 1740.7-2942 (Goldwurm et
al. 1992, Bouchet et al. 1991).  Since the gamma-ray line luminosity
is independent of both of the uncertainties mentioned above (namely
depletion time scale and $F_{esc}$), we are in a position to estimate
the luminosity fairly accurately (sec. 2.5).  We find that, even under
the most optimistic of circumstances, the calculated flux is much
below the detection thresholds of present instruments (though perhaps
detectable by the Ge spectrometer planned for the INTEGRAL mission).
Thus, at least within the ADAF paradigm, we can rule out the
suggestion of Martin et al.  However, as we have mentioned earlier, Li
could conceivably be produced during SXT outbursts by a different
process for which our ADAF-based estimates may not be applicable.

The abundance of Li in Pop I
material in the Galaxy is about $\log(n_{Li}/n_H)\sim-9$, whereas the
primordial abundance as measured in halo stars is significantly lower,
$\log(n_{Li}/n_H)\sim-9.8$ (Boesgaard \& Steigman 1985). 
What is the origin of the
extra Li in the Galactic disk?  Cosmic ray spallation alone cannot
explain the observations, and it is proposed that there needs to be an
anomalous component of low energy cosmic rays with energy $\sim$ tens
of MeV per nucleon (e.g. Reeves et al. 1990).  
ADAFs provide precisely the kind of
environment and particle energies needed by the observations and it is
interesting to ask if Li production in accreting black holes or
neutron stars could be a significant source of the Li observed in the
Galaxy.

If we take the total baryonic mass of the Galactic disk to be
$\sim10^{11}M_\odot$, then the mass of Li is $\sim7\times10^{-9}\times
0.75\times10^{-11}M_\odot\sim500M_\odot$.  The rate at which Li is
ejected into the ISM by an accreting black hole is given by equation
(2-11) multiplied by the escape fraction $F_{esc}$, i.e.
\begin{equation}
\dot M_{Li}=2.47\times10^{-10}F_{esc}{m\dot m^2\over\alpha^2}
M_\odot\,{\rm yr^{-1}}.
\end{equation}
Even by using quite optimistic estimates of the various parameters, we
find that it is very hard to produce $500M_\odot$ of Li during the
life of the Galaxy from known accreting systems.  We consider four
cases:

(i) For black hole SXTs, we have $m\sim10$, $\dot m\sim0.003$,
$\alpha\sim0.3$.  If we assume that the integrated Li production from
outbursts is 10 times larger than during quiescence
(cf. eq. 3-16), and if we take a
large $F_{esc}\sim0.1$, then we still obtain only $\sim
2.5\times10^{-13}M_\odot{\rm yr^{-1}}$ of Li from a single SXT.  The Galaxy
needs to have a steady population of $\sim2\times10^5$ active SXTs
over the lifetime of the Galaxy to produce the observed Li.  Current
estimates of the SXT population are, however, only $\sim {\rm
few}\times10^3$ active systems (Tanaka \& Shibazaki 1996 and references 
therein).

(ii) Consider Cyg X-1 like objects: $m\sim10$, $\dot m\sim0.1$,
$\alpha\sim0.3$.  Again, taking $F_{esc}\sim0.1$, we need about 2000
active objects at any given time to produce the required Li.  The
number of bright X-ray binaries in the Galaxy is only $\sim10^2$, and
only a small fraction of these are Cyg X-1 like objects.

(iii) For neutron star SXTs, we optimistically
assume that most of the systems have $r_A<30$ in quiescence and are
therefore as efficient as the black hole systems at producing Li.
Setting $m=1.4$, $\dot m\sim0.003$ (optimistic), $\alpha\sim0.3$,
$F_{esc}\sim1$ (assuming a propeller), each system ejects
$\sim3.5\times10^{-14} M_\odot{\rm yr^{-1}}$ of Li into the ISM.  We need a
steady population of over $10^6$ active objects to explain the
observed Li in the Galactic disk, which seems unlikely.

(iv) Finally, we consider the black hole at the Galactic Center.
We have shown in sec 3.5 that
the current level of activity would produce very little $^7$Li
during the age of the Galaxy $\sim 10^{10}$ yr.
Li production at the Galactic Center becomes interesting only if
the Galaxy has experienced an active accretion episode in the past, similar
to that in Seyfert galaxies. Let us assume that the present mass of the 
the black hole ($m\sim 2.5\times 10^6$) was accumulated primarily
during episodes of ${\dot m}\sim 0.1$ lasting for a total duration
of $\sim 4.5\times 10^8$yr. Then equation
(4-1) suggests that such a source would have ejected 
$\sim 3\times 10^3M_{\sun}$ 
of $^7$Li for $\alpha=0.3$ and $F_{esc}=0.1$, which is more than enough 
to account for the total mass of Li in the Galactic disk. 

We thus conclude that none of the known populations of X-ray binaries in the
Galaxy is capable of supplying the observed Li in Pop
I stars. A period of active accretion in the past at the Galactic Center 
could have ejected sufficient Li, but it is unclear if the Li would have spread
over the entire disk.

\acknowledgments We thank John Bahcall and Jeff McClintock for useful 
discussions and suggestions.  This work was supported in part by NASA
grant NAG 5-2837.  I. Y. acknowledges the support of SUAM foundation.

\clearpage


\begin{references}
\reference{A} Abramowicz, M. A., Chen, X., Kato, S., Lasota, J. P., \& Regev,
O. 1995, ApJ, 438, L37
\reference{A} Asai, K. et al. 1996, PASJ, 48, 257
\reference{A} Barret, D., McClintock, J. E., \& Grindlay, J. E. 1996, ApJ, 
473, 963
\reference{A} Bodansky, D. Jacobs, W. W., \& Oberg, D. L. 1975, ApJ, 202, 222
\reference{A} Boesgaard, A. M. \& Steigman, G. 1985, ARA\&A, 23, 319
\reference{A} Brown, J. A., Sneden, C., Lambert, D. L., \& Dutchover, E., Jr. 
1989, ApJS, 71, 293
\reference{A} Browne, E. \& Firestone, R. B. 1986, Table of Radioactive Isotopes,
New York: Wiley-Interscience Publishers
\reference{A} Bouchet, L. et al. 1991, ApJ, 383, L45
\reference{A} Burcham, W. E. et al. 1958, Nucl. Phys., 5, 141
\reference{A} Chen, X., Abramowicz, M. A., \& Lasota, J.-P. 1997, ApJ, 
476, in press
\reference{A} Chen, X., Abramowicz, M., Lasota, J. P., Narayan, R., \& Yi, I. 
1995, ApJ, 443, L61
\reference{A} Cox, J. P. \& Giuli, R. T. 1968, Principles of Stellar Structure, 
New York: Gordon \& Breach Science Publishers
\reference{A} Dearborn, D. S. P., Schramm, D. N., Steigman, G., \& Truran, J.
1989, ApJ, 347, 455
\reference{A} Duncan,  1981, ApJ, 248, 651
\reference{A} Eggleton, P. P. 1983, ApJ, 268, 368
\reference{A} Esin, A. A. 1996, ApJ, in press
\reference{A} Field, G. B. \& Rogers, R. D. 1993, ApJ, 403, 94
\reference{A} Filippenko, A. V., Matheson, T., \& Barth, A. J. 1995, ApJ, 455, 
L139
\reference{A} Foster, R. S. et al. 1996, ApJ, 467, L81
\reference{A} Frank, J., King, A. R., and Raine, D. 1992, Accretion 
Power in Astrophysics, Cambridge: Cambridge University Press
\reference{A} Garcia-Lopez, R. J., Rebolo, R. \& Martin, E. L. 1994, A\&A, 282, 
518
\reference{A} Goldwurm, A. et al. 1992, ApJ, 389, L79
\reference{A} Harlaftis, E. T., Horne, K., \& Filippenko, A. V. 1996, PASP, 
108, 762
\reference{A} Hjellming, R. M., Calovini, T., Han, X.-H., C{\'o}rdova, F. A.
1988, ApJ, 335, L75
\reference{A} Hjellming R. M. \& Han, X.-H. 1995, in X-Ray Binaries ed. W. H. G. 
Lewin, J. van Paradijs, \& E. P. J. van den Heuvel 
(Cambridge: Cambridge University Press), p308
\reference{A} Illarionov, A. F. \& Sunyaev, R. A. 1975, A\&A, 39, 185
\reference{A} Jin, L. 1990, ApJ, 356, 501
\reference{A} Kozlovsky, B. \& Ramaty, R. 1974, ApJ, 191, L43
\reference{A} Kulkarni, S. R. \& Hester, J. J. 1988, Nature, 335, 801
\reference{A} Martin, E. L., Casares, J., Charles, P. A., \& Rebolo, R. 1995, 
A\&A, 303, 785
\reference{A} Martin, E. L., Casares, J., Molaro, P., Rebolo, R., \& Charles, P. 
A.
1996, New Astron., 1, 197
\reference{A} Martin, E. L., Magazzu, A., \& Rebolo, R. 1992a, A\&A, 257, 186
\reference{A} Martin, E. L., Rebolo, R. Casares, J., \& Charles, P. A. 1992b, 
Nature, 358, 129
\reference{A} Martin, E. L., Rebolo, R., Casares, J., \& Charles, P. A. 1994a, 
ApJ, 435, 791
\reference{A} Martin, E. L., Spruit, H. C., \& van Paradijs, J. 1994b, A\&A, 291, 
L43
\reference{A} McClintock, J. E. \& Remillard, R. A. 1990, ApJ, 350, 386
\reference{A} Meneguzzi, M., Audouze, J., \& Reeves, H. 1971, A\&A, 15, 337
\reference{A} Meneguzzi, M. \& Reeves, H. 1975, A\&A, 40, 99
\reference{A} Mitsuda, K., Asai, K., Vaughan, B., \& Tanaka, Y. 1996, in Proc.  of 
X-Ray Imaging and Spectroscopy of Cosmic Hot Plasmas, International Symposium
on X-ray Astronomy (Waseda Univ., Tokyo March 11-14, 1996), in press
\reference{A} Narayan, R. 1996, ApJ, 462, 136
\reference{A} Narayan, R., Barret, D., \& McClintock, J. E. 1997, ApJ, 
482, in press
\reference{A} Narayan, R., Kato, S., \& Honma, F. 1997, ApJ, 476, in press
\reference{A} Narayan, R. \& Yi, I. 1994, ApJ, 428, L13
\reference{A} Narayan, R. \& Yi, I. 1995a, ApJ, 444, 231
\reference{A} Narayan, R. \& Yi, I. 1995b, ApJ, 452, 710
\reference{A} Narayan, R., Yi, I., \& Mahadevan, R. 1995, Nature, 374, 623
\reference{A} Narayan, R., McClintock, J. E., \& Yi, I. 1996, ApJ, 457, 821
\reference{A} Pallavicini, G., Randich, S., \& Giampapa, M. 1992, A\&A, 253, 185
\reference{A} Pilachowski, C. A., Mould, J. R., \&  Seigel, M. J. 1984, ApJ, 
282, L17
\reference{A} Pinsonneault, M. H., Deliyannis, C. P., \& Demarque, P. 1992, 
ApJS, 78, 179
\reference{A} Ramadurai, S. \& Rees, M. J. 1985, MNRAS, 215, 53p
\reference{A} Rees, M. J., Begelman, M. C., Blandford, R. D., Phinney, E. S.
1982, Nature, 295, 17
\reference{A} Reeves, H. 1974, ARA\&A, 12, 437
\reference{A} Reeves, H., Richer, J., Sato, K., \& Terasawa, N. 1990, ApJ, 355, 18
\reference{A} Rytler, C., Reeves, H., Gradsztajn, E., \& Audouze, J. 1970, A\&A,
8, 389
\reference{A} Shahbaz, T. et al. 1994, MNRAS, 271, L10
\reference{A} Shapiro, S. L., Lightman, A. P., \& Eardley, D. M. 1976,
ApJ, 204, 187
\reference{A} Spruit, H., Matsuda, T., Inoue, M., \& Sawada, K. 1987, MNRAS, 229,
517
\reference{A} Tanaka, Y. \& Shibazaki, N. 1996, ARA\&A, 34, 607
\reference{A} Thorburn, J. A., Hobbs, L. M., Deliyannis, C. P., Pinsonneault,
M. H. 1993, ApJ, 415, 150
\reference{A} Woosley, S. E. 1993, ApJ, 405, 273
\reference{A} Zahn, J. P. 1994, A\&A, 288, 829
\reference{A} Zahn, J. P. \& Bouchet, L. 1989, A\&A, 223, 112
\end{references}
\end{document}